\pdfoutput=1
\documentclass[letterpaper]{article}

\usepackage{aaai}
\usepackage{times}
\usepackage{helvet}
\usepackage{courier}
\usepackage{graphicx} 
\usepackage{subfig} 

\usepackage{algorithm}
\usepackage{algorithmic}

\renewcommand{\footnoterule}{%
  \kern -3pt
  \hrule width 2in
  \kern 2.6pt
}
\nocopyright

\usepackage{amsmath, amsfonts, amsthm}
\usepackage{multirow}
\usepackage{listings}
\usepackage{placeins}

\theoremstyle{definition}
\newtheorem{definition}{Definition}

\newtheorem{example}{Example}

\newcommand{\npcite}[1]{\citeauthor{#1} \citeyear{#1}}
\newcommand{\citet}[1]{\citeauthor{#1} \shortcite{#1}}
\newcommand{\citep}[2]{(#1 \citeauthor{#2} \citeyear{#2})}

\begin{document}

\title{Learning Tractable Probabilistic Models for Fault Localization}
\author{{\bf Aniruddh Nath}
\and
{\bf Pedro Domingos} \\
Department of Computer Science \& Engineering \\
University of Washington \\
Seattle, WA 98195, U.S.A. \\
{\it \{nath, pedrod\}@cs.washington.edu}
}
\maketitle
\begin{abstract}
  In recent years, several probabilistic techniques have been applied
to various debugging problems. However, most existing probabilistic debugging
systems use relatively simple statistical models, and fail to generalize
across multiple programs.
In this work, we propose {\em Tractable Fault Localization Models}
(TFLMs) that can be learned from data, and probabilistically infer
the location of the bug.
While most previous statistical debugging methods generalize over many
executions of a single program, TFLMs are trained on a corpus of previously seen
buggy programs, and learn to identify recurring patterns of bugs.
Widely-used fault localization techniques such as {\sc Tarantula} evaluate the
suspiciousness of each line in isolation; in contrast, a TFLM defines a
joint probability distribution over buggy indicator variables for each line.
Joint distributions with rich dependency structure are often computationally
intractable; TFLMs avoid this by exploiting recent developments in tractable
probabilistic models (specifically, Relational SPNs).
Further, TFLMs can incorporate additional sources of
information, including coverage-based features such as {\sc Tarantula}.
We evaluate the fault localization performance of TFLMs that
include {\sc Tarantula} scores as features in the probabilistic model. Our
study shows that the learned TFLMs isolate bugs more effectively than 
previous statistical methods or using {\sc Tarantula} directly.

\end{abstract}

\section{Introduction}
\label{sec:intro}
According to a 2002 NIST study \cite{nist02}, software bugs cost the US
economy an estimated \$59.5 billion per year. While some of these costs are
unavoidable, the report claimed that an estimated \$22.2 billion could be saved
with more effective tools for the identification and removal of software
errors. Several other sources estimate that over 50\% of software development
costs are spent on debugging and testing \cite{hailpern&santhanam02}.

The need for better debugging tools has long been recognized. The goal of
automating various debugging tasks has motivated a large body of research in
the software engineering community. However, this line of work has only recently
begun to take advantage of recent advances in probabilistic models, and their
inference and learning algorithms.

In this work, we apply state-of-the-art probabilistic methods to the problem of
fault localization. We propose {\em Tractable Fault Localization Models}
(TFLMs)
that can be learned from a corpus of known buggy programs (with the bug
locations annotated). The trained model can then be used to infer the probable
locations of buggy lines in a previously unseen program.
Conceptually, a TFLM is a probability distribution over programs in a
given language, modeled jointly with any attributes of interest (such as bug
location indicator variables, or diagnostic features).
Conditioned on a specific program, a TFLM defines a joint probability
distribution over the attributes.

The key advantage of probabilistic models is their ability to learn from
experience. 
Many software faults are instances of a few common error patterns, such as
off-by-one errors and use of uninitialized values \cite{brun&ernst04}.
Human debuggers improve with experience as they encounter
more of these common fault patterns, and learn to recognize them in new
programs. Automated debugging systems should be able to do the same.

Another advantage of probabilistic models is that they allow multiple sources of
information to be combined in a principled manner. The relative contribution of
each feature determined by its predictive value in the training corpus, rather
than by a human expert. A TFLM can incorporate as features the outputs of
other fault localization systems, such as the {\sc Tarantula} hue
\cite{jones&al02} of each line.

In recent years, there has been renewed interest in learning rich, tractable
models, on which exact probabilistic inference can be performed in polynomial
time (e.g.\ Sum-Product Networks; \npcite{poon&domingos11}).
TFLMs build on Relational Sum-Product Networks \cite{nath&domingos15} to enable
exact inference in space and time linear in the size of the program.
We empirically compare TFLMs to the widely-used {\sc Tarantula} fault
localization method, as well as the Statistical Bug Isolation (SBI) system,
on four mid-sized C programs. TFLMs outperform the other
systems on three of the four test subjects.


\section{Background}
\subsection{Coverage-based Fault Localization}
\label{sec:spectra}
Coverage-based debugging methods isolate the bug's location by analyzing the
program's coverage spectrum on a set of test inputs.
These approaches take the following as input:
\begin{enumerate}
\item a set of unit tests;
\item a record of whether or not the program passed each test;
\item program traces, indicating which components (usually lines) of the
program were executed when running each unit test.
\end{enumerate}
Using this information, these methods produce a suspiciousness
score for each component in the program. The most well-known method in this
class is the {\sc Tarantula} system \cite{jones&al02}, which uses the following
scoring function:
\begin{equation*}
  S_{\mathit{Tarantula}}(s) = 
\frac{\frac{\mathit{Failed}(s)}{\mathit{TotalFailed}}}
{
  \frac{\mathit{Passed}(s)}{\mathit{TotalPassed}}
  +
  \frac{\mathit{Failed}(s)}{\mathit{TotalFailed}}
}
\end{equation*}
Here, $\mathit{Passed}(s)$ and $\mathit{Failed}(s)$ are respectively the number of passing and
failing test cases that include statement $s$, and $\mathit{TotalPassed}$ and
$\mathit{TotalFailed}$ are the number of passing and failing test cases respectively.
In an empirical evaluation \cite{jones&harrold05}, {\sc Tarantula} was shown to
outperform previous methods such as cause transitions \cite{cleve&zeller05},
set union, set intersection and nearest neighbor \cite{renieris&reiss03}, making
it the state of the art in fault localization at the time.
Since the publication of that experiment, a few other scoring functions have
been shown to outperform {\sc Tarantula} under certain conditions
\cite{abreu&al09}.
Nonetheless, {\sc Tarantula} remains the most well-known method in this class.

\subsection{Probabilistic Debugging Methods}
\label{sec:statdebugging}
\subsubsection{Per-Program Learning}
\label{sec:statdebugging_pp}
Several approaches to fault localization make use of statistical and
probabilistic methods. Liblit et al. proposed several influential statistical
debugging methods. Their initial approach \cite{liblit&al03} used
$\ell_1$-regularized logistic regression to predict non-deterministic program
failures. (The instances are runs of a program, the features are instrumented
program predicates, and the models are trained to predict a binary `failure'
variable. The learned weights of the features indicate which predicates are the
most predictive of failure.)
In later work \cite{liblit&al05}, they use a likelihood ratio
hypothesis test to determine which {\em predicates} (e.g.\ branches, sign of
return value) in an instrumented program are predictive of program failure.
Zhang et al. \shortcite{zhang&al11} evaluate several other hypothesis testing
methods in a similar setting.

The SOBER system \cite{liu&al05,liu&al06} improves on Liblit et al.'s 2005
approach by taking into account the fact that a program predicate can be
evaluated multiple times in a single test case. They learn conditional
distributions over the probability of a predicate evaluating to {\tt true},
conditioned on the success or failure of the test case. When these conditional
distributions differ (according a statistical hypothesis test), the predicate is
considered to be `relevant' to the bug. The HOLMES system \cite{chilimbi&al09}
extends Liblit et al.'s approach along another direction, analyzing path
profiles instead of instrumented predicates.

Wong et al. \shortcite{wong&al08,wong&al12} use a crosstab-based statistical analysis
to quantify the dependence between statement coverage and program failure. Their
approach can be seen as a hybrid between the Liblit-style statistical analysis
and {\sc Tarantula}-style spectrum-based analysis.
Wong et al.\ \shortcite{wong&qi09,wong&al12b} also proposed two neural network-based fault localization
techniques trained on program traces.
Ascari et al.
\shortcite{ascari&al09} investigate the use of SVMs in a similar setting.

Many of the methods described above operate under the
assumption that the program contains exactly one bug. Some of these techniques
have nevertheless been evaluated on programs with multiple faults, using an
iterative process where the bugs are isolated one by one.
Briand et al. \shortcite{briand&al07} explicitly extend {\sc Tarantula} to the
multiple-bug case, by learning a decision tree to partition failing test cases.
Each partition is assumed to model a different bug.
Statements are ranked by suspiciousness using a {\sc Tarantula}-like
scoring function, with the scores computed separately for each partition.
Other clustering methods have also been applied to test cases; for example,
Andrzejewski et al. \shortcite{andrzejewski&al07} use a form of LDA
to discover latent `bug topics'.

%
%
%

\subsubsection{Generalizing Across Programs}
\label{sec:statdebugging_ap}
The key limitation of the statistical and machine learning-based approaches
discussed above is that they only generalize over many executions of a single
program. Ideally, a machine learning-based debugging system should be able to
generalize over multiple programs (or, at least, multiple sequential versions of
a program).
As discussed above, many software defects are instances of
frequently occurring fault patterns; in principle, a machine learning model
can be trained to recognize these patterns and use them to more effectively
localize faults in new programs.

This line of reasoning has received relatively little attention in the automated
debugging literature.
The most prominent approach
is the Fault Invariant Classifier (FIC) of Brun and Ernst \shortcite{brun&ernst04}.
FIC is not a fault localization algorithm in the sense of
{\sc Tarantula} and the other approaches discussed above. Instead of localizing
the error to a particular line, FIC outputs {\em fault-revealing properties}
that can guide a human debugger to the underlying error. These properties can be
computed using static or dynamic program analysis; FIC uses the Daikon dynamic
invariant detector \cite{ernst&al01}. At training time, properties are computed
for pairs of buggy and fixed programs; properties that occur in the buggy
programs but not the fixed programs are labeled as `fault-revealing'. The
properties are converted into program-independent feature vectors, and an SVM or
decision tree is trained to classify properties as fault-revealing or
non-fault-revealing. The trained classifier is then applied to properties
extracted from a previously unseen, potentially faulty program, to reveal
properties that indicate latent errors.

\subsection{Tractable Probabilistic Models}
\label{sec:spn}
\subsection{Sum-Product Networks}

A sum-product network (SPN) is a rooted directed acyclic graph with univariate
distributions at the leaves; the internal nodes are (weighted) sums and
(unweighted) products.
\begin{definition} \cite{gens&domingos13}
\begin{enumerate}
\item A tractable univariate distribution is an SPN.
\item A product of SPNs with disjoint scopes is an SPN. (The scope of an SPN is the set of variables that appear in it.)
\item A weighted sum of SPNs with the same scope is an SPN, provided all weights are positive.
\item Nothing else is an SPN.
\end{enumerate}
\end{definition}
%
%
\begin{figure}
  \centering
  \includegraphics[width=3.3in]{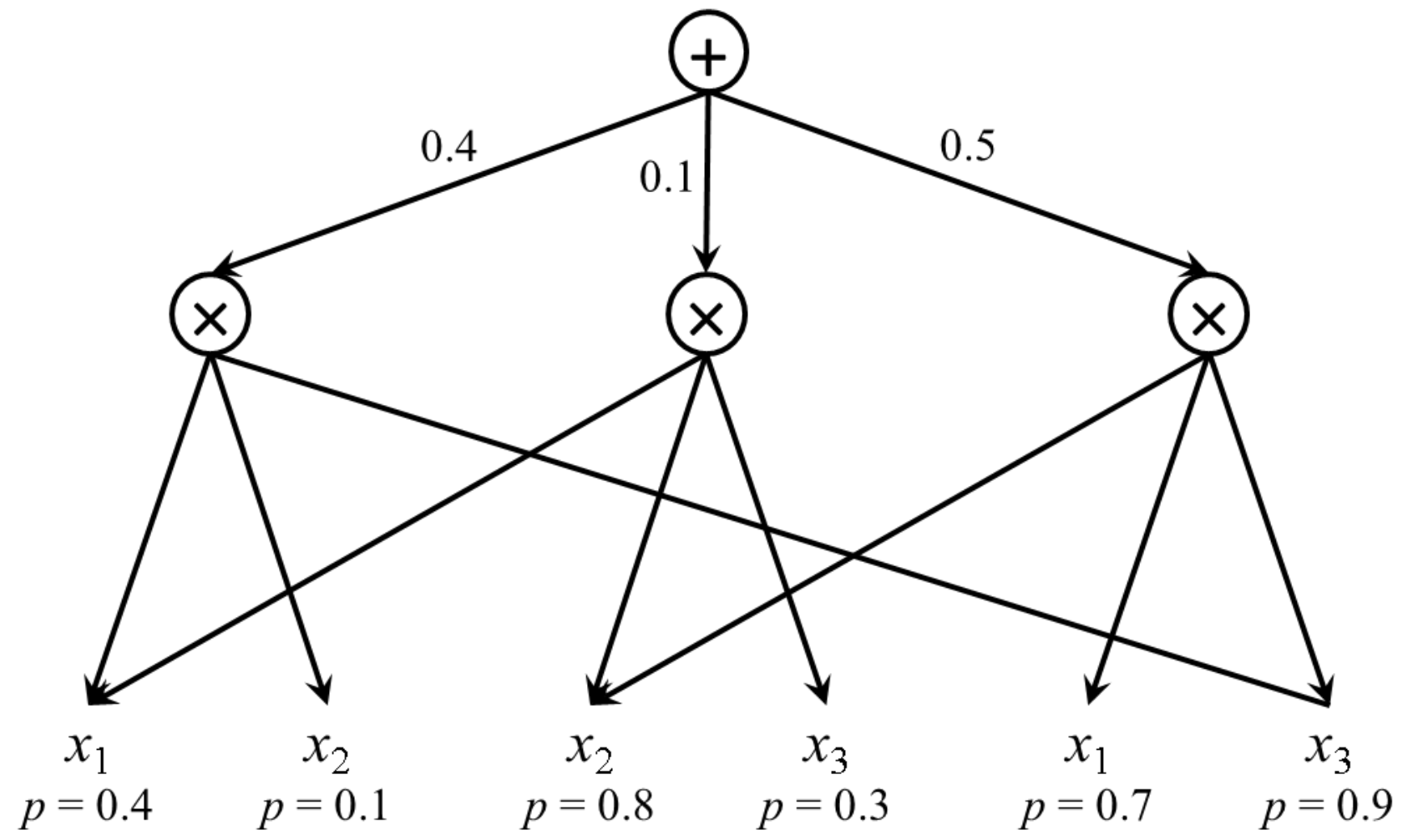}
  \caption{Example SPN over the variables $x_1$, $x_2$ and $x_3$.
    All leaves are Bernoulli distributions, with the given parameters.
    The weights of the sum node are indicated next to the corresponding edges.}
  \label{fig:eg_spn}
\end{figure}
Intuitively, an SPN (fig.~\ref{fig:eg_spn})
can be thought of as an alternating set of mixtures (sums)
and decompositions (products) of the leaf variables. If the values at the leaf
nodes are set to the partition functions of the corresponding univariate
distributions, then the value at the root is the partition function (i.e. the sum
of the unnormalized probabilities of all possible assignments to the leaf
variables). This allows the partition function to be computed in time linear in
the size of the SPN.

If the values of some variables are known, the leaves corresponding to
those variables' distributions are set to those values' probabilities,
and the remainder are replaced by their (univariate) partition
functions.
This yields the unnormalized probability of the evidence, which can be divided
by the partition function to obtain the normalized probability.
The most probable state of the SPN, viewing sums as marginalized-out
hidden variables, can also be computed in linear time.
The first learning algorithms for sum-product networks used a fixed network
structure, and only optimized the weights
\cite{poon&domingos11,amer&todorovic12,gens&domingos12}.
More recently, several structure learning algorithms for SPNs have also been
proposed \cite{dennis&ventura12,gens&domingos13,peharz&al13}.

\subsection{Relational Sum-Product Networks}

SPNs are a propositional representation, modeling instances as independent and
identically distributed (i.i.d.). Although the i.i.d. assumption is widely used
in statistical machine learning, it is often an unrealistic assumption.
In practice, objects usually interact with each other; Statistical Relational
Learning algorithms can capture dependencies between
objects, and make predictions about relationships between them.

Relational Sum-Product Networks (RSPNs; \npcite{nath&domingos15})
generalize SPNs by modeling a set of instances jointly, allowing them to
influence each other's probability distributions, as well as modeling
probabilities of relations between objects.
RSPNs can be seen as templates for constructing SPNs, much like Markov Logic
Networks (\npcite{richardson&domingos06}) are templates for Markov networks.
RSPNs also require as input a part decomposition, which describes the part-of
relationships among the objects in the mega-example.
Unlike previous high-treewidth tractable relational models such as TML
\cite{domingos&webb12}, RSPNs can generalize across mega-examples of varying
size and structure.

\section{Tractable Fault Localization}
\subsection{Tractable Fault Localization Models}

A {\em Tractable Fault Localization Model} (TFLM) defines a probability
distribution over programs in some deterministic language $L$.
The distribution may
also model additional variables of interest that are not part of the program
itself; we refer to such variables as {\em attributes}.
In the fault localization setting, the
important attribute is a $\textit{buggy}$ indicator variable on each line.
Other informative features may also be included as attributes; for
instance, one or more coverage-based metrics may be included for each line.

\noindent More formally, consider a language whose grammar $L = (V, \Sigma, R,
S)$ consists of:
\begin{itemize}
  \item $V$ is a set of {\em non-terminal} symbols;
  \item $\Sigma$ is a set of {\em terminal} symbols;
  \item $R$ is a set of production rules of the form $\alpha \rightarrow \beta$,
    where $\alpha \in V$ and $\beta$ is a string of symbols in $V \cup \Sigma$;
  \item $S \in V$ is the start symbol.
\end{itemize}
\begin{definition}
  A {\em Tractable Fault Localization Model} for language $L$ consists of:
  \begin{itemize}
    \item a map from non-terminals in $V$ to sets of {\em attribute
      variables} (discrete or continuous);
    \item for each symbol $\alpha \in V$, a set of latent subclasses
          $\alpha_1, \ldots, \alpha_k$;
    \item $\pi_S$, a probability distribution over subclasses of the start symbol $S$;
    \item for each each subclass $\alpha_i$ of $\alpha$,
    \begin{itemize}
      \item a univariate distribution $\psi_{\alpha_i,x}$ over each attribute
        $x$ associated with $\alpha$;
      \item for each rule $\alpha \rightarrow \beta$ in $L$,
        a probability distribution $\rho_{\alpha_i}$ over rules $\alpha_i \rightarrow \beta$;
      \item for each rule $\alpha_i \rightarrow \beta$, for each non-terminal
        $\alpha^\prime \in \beta$, a distribution
        $\pi_{(\alpha_i \rightarrow \beta),\alpha^\prime}$ over subclasses of $\alpha^\prime$.
    \end{itemize}
  \end{itemize}
\end{definition}
The univariate distribution over each attribute may be replaced with a joint
distribution over all attributes, such as a logistic regression model within
each subclass that predicts the value of the $\textit{buggy}$ attribute, using one or
more other attributes as features. However, for simplicity, we present the
remainder of this section with the attributes modeled as a product of
univariate distributions, and assume that the attributes are discrete.

%

Being defined over the grammar of the programming language, TFLMs can capture
information extremely useful for the fault localization task. For example, a
TFLM can represent different fault probabilities for different symbols in the
grammar. In addition, the latent subclasses give TFLMs a degree of
context-sensitivity; the same symbol can be more or less likely to contain a
fault depending on its latent subclass, which is probabilistically dependent on
the subclasses of ancestor and descendent symbols in the parse tree. This
makes TFLMs much richer than models like logistic regression, where the features
are independent conditioned on the class variable.
Despite this representational power, exact inference in TFLMs is still
computationally efficient.

\begin{example}
The following rules are a fragment of the grammar of a Python-like
language:
\begin{lstlisting}[mathescape,basicstyle=\footnotesize]
  while_stmt $\rightarrow$ 'while' condition ':' suite
  condition $\rightarrow$ expr operator expr
  condition $\rightarrow$ 'not' condition
\end{lstlisting}
We refer to the above rules as $r_1$, $r_2$ and $r_3$ respectively.

The following is a partial specification of a TFLM over this grammar, with the
{\tt while\_stmt} symbol as root.
\begin{itemize}
  \item All non-terminal symbols have $\textit{buggy}$ and
    $\textit{suspiciousness}$ attributes. $\textit{buggy}$ is a fault indicator, and
    $\textit{suspiciousness}$ is a diagnostic attribute, such as a {\sc Tarantula} score.
  \item
    Each non-terminal has two latent subclass symbols.
    For example, {\tt while\_stmt} has subclasses ${\tt while\_stmt}_1$
    and ${\tt while\_stmt}_2$.
  \item The distribution over start symbols is:
      $\pi({\tt while\_stmt}_1) = 0.4$, \quad
      $\pi({\tt while\_stmt}_2) = 0.6$.
  \item For subclass symbol ${\tt while\_stmt}_1$ (subclass subscripts omitted):
  \begin{itemize}
    \item $\psi_{buggy} \sim Bernoulli(0.01)$
    \item $\psi_{suspiciousness} \sim \mathcal{N}(0.4,0.05)$
    \item $\rho(r_1) = 1.0$, since {\tt while\_stmt} has a single rule.
    \item The distributions over child symbol subclasses for $r_1$ are:
        $\pi_{r_1, {\tt condition}}({\tt condition}_1) = 0.7$, \quad
        $\pi_{r_1, {\tt condition}}({\tt condition}_2) = 0.3$, \quad
        $\pi_{r_1, {\tt suite}}({\tt suite}_1) = 0.2$, \quad
        $\pi_{r_1, {\tt suite}}({\tt suite}_2) = 0.8$.
  \end{itemize}
\end{itemize}
(The complete TFLM specification would have similar definitions for all the
other subclass symbols in the model.)
\end{example}

\subsubsection{Semantics}

Conceptually, a TFLM defines a probability distribution over
all programs in $L$, and their attributes. More formally, the
joint distribution $P(T,A,C)$ is defined over:
\begin{itemize}
  \item a parse tree $T$;
  \item an attribute assignment $A$, specifying values of all attributes of all
    non-terminal symbols in $T$;
  \item latent subclass assignment $C$ for each non-terminal in $T$.
\end{itemize}

For parse tree $T$ containing rules $r_1 = \alpha_1 \rightarrow \beta_1, r_2 = \alpha_2
\rightarrow \beta_2, \ldots, r_n = \alpha_n \rightarrow \beta_n$, and root symbol
$\alpha_R$,
$
  P(T,A,C) = \prod_{r_i} \bigg(
                \rho_{C(\alpha_i)}(\alpha_i \rightarrow \beta_i)
            \times \prod_{\alpha^\prime \in \beta_i}
  \pi_{(C(\alpha_i)\rightarrow\beta_i),\alpha^\prime}(C(\alpha^\prime))
            \times \prod_{x \in attr(\alpha_i)} \psi_{C(\alpha_i),x}(A(x))
                \bigg)
            \times \pi_S(C(\alpha_R))
$

\subsubsection{Inference}
Like RSPNs, inference in TFLMs is performed by grounding out the model into an
SPN. 
The SPN is constructed in a recursive top-down manner, beginning
with the start node:
\begin{itemize}
  \item
    Emit a sum node over subclasses for the started node, weighted
    according to $\pi_S$.
    Let the current symbol $\alpha$ be the start symbol, and let $\alpha_i$ be
    its subclass.
  \item
    Emit a product node with one child for each attribute of $\alpha$, and a
    child for the subprogram rooted at $\alpha$.
  \item
    For each attribute $x$ of $\alpha$, emit a sum node over the attribute
    values for the current symbol, weighted by $\psi_{\alpha_i,x}$.
  \item
    Emit a sum node over production rules $\alpha \rightarrow \beta$ for the current
    symbol, according to $\rho_{\alpha_i}$. (Note that when grounding a TFLM
    over a known parse tree, all but one child of this sum node is zeroed out,
    and need not be grounded.)
  \item Recurse over each non-terminal $\alpha^\prime$ in $\beta$, choosing its
    subclass via a sum node weighted by $\pi_{(\alpha_i \rightarrow \beta),
    \alpha^\prime}$.
\end{itemize}

\subsubsection{Learning}
\label{sec:learning}

The learning problem in TFLMs is to estimate the $\pi$ and $\psi$ distributions
from a training corpus of programs, with known attribute values but unknown
latent subclasses.
(The $\rho$ distributions
have no effect on the distribution of interest, since we assume that every
program can be unambiguously mapped to a parse tree.)

As is commonly done with SPNs, we train the model via hard EM.
In the E-step, given the current parameters of $\pi$ and $\psi$, we compute
the MAP state of the training programs (i.e.\ the latent subclass assignment
that maximizes the log-probability).
In the M-step, we
re-estimate the parameters of $\pi$ and $\psi$, choosing
the values that maximize the log-probability.
These two steps are repeated until convergence, or for a fixed number of
iterations.

If the attributes are modeled jointly rather than as a product of univariate
distributions, retraining the joint model in each iteration of EM is
computationally expensive. A more efficient alternative is to use a product of
univariates during EM, in order to learn a good subclass assignment. The joint
model is then only trained once, at the conclusion of EM.

\section{Experiments}
We performed an experiment to determine whether TFLM's ability to combine a
coverage-based fault localization system with learned bug patterns improves
fault localization performance, relative to using the coverage-based system
directly. As a representative coverage-based method, our study used {\sc
Tarantula}, one of the most widely-used approaches in this class, and a common
comparison system for fault localization algorithms.
We also compared to the statement-based version of Liblit et al.'s Statistical
Bug Isolation (SBI) system \cite{liblit&al05}, as adapted by Yu et al.
\shortcite{yu&al08}.
SBI serves as a representative example of a lightweight statistical method for
fault localization.

\subsection{Subjects}
\begin{table}
\centering
\caption{Subject programs}
\begin{tabular}{ccccccc}
\hline
Program
& Versions
& Executable LOC
& Buggy vers.
\\ \hline
grep &   4   & 3368 $\pm$ 122 & 8  $\pm$ 5 \\ 
gzip &   5   & 1905 $\pm$ 124 & 7  $\pm$ 3 \\ 
flex &   5   & 3907 $\pm$ 254 & 10 $\pm$ 4 \\ 
sed  &   7   & 2154 $\pm$ 389 & 3  $\pm$ 1 \\ 
\hline
\end{tabular}
\label{tab:datasets}
\end{table}


We evaluated TFLMs on four mid-sized C programs (table~\ref{tab:datasets}) from the
Software-artifact Infrastructure Repository (\npcite{sir}).
All four test subjects are real-world programs, commonly used to evaluate fault
localization approaches.
The repository contained several sequential versions of each program, each
with several buggy versions. The repository also contained a suite of between
124 and 525 TSL tests for each version, which we used to compute the {\sc
Tarantula} scores.

The number of executable lines was measured by the gcov tool.
We excluded buggy versions where the bug occurred in a non-executable line
(e.g.\ lines excluded by preprocessor directives), or
consisted of line insertions or deletions.
Unlike most previous fault localization studies that use these
subjects, we do not exclude versions for which the test results were uniform
(i.e.\ consisting entirely of passing or failing tests). Although coverage-only
methods such as {\sc Tarantula} can provide no useful information in the case of
uniform test suites, TFLMs can still make use of learned contextual information
to determine that some lines are more likely than others to contain a fault.

\subsection{Methodology}
\begin{table}
\centering
\caption{Localization accuracy (fraction of lines skipped)}
\begin{tabular}{cccc}
\hline
Program
& TFLM
& Tarantula
& SBI
\\ \hline
grep
& {\bf 0.645}
& 0.640
& 0.564
\\
gzip
& 0.516
& {\bf 0.682}
& 0.540
\\
flex
& {\bf 0.770}
& 0.704
& 0.618
\\
sed
& {\bf 0.927}
& 0.851
& 0.603
\\ \hline
\end{tabular}
\label{tab:results}
\end{table}

\begin{table}
\centering
\caption{TFLM (with Tarantula feature) vs Tarantula alone}
\begin{tabular}{cccc}
\hline
Program
& TFLM wins
& Ties
& Tarantula wins
\\ \hline
grep
& {\bf 18}
& 0
& 14
\\
gzip
& 11
& 0
& {\bf 23}
\\
flex
& {\bf 31}
& 0
& 18
\\
sed
& {\bf 17}
& 0
& 7
\\ \hline
\end{tabular}
\label{tab:compare}
\end{table}

\begin{table}
\centering
\caption{TFLM learning and inference times}
\begin{tabular}{ccc}
\hline
Program
& Avg. learn time (s)
& Avg. infer time (s)
\\ \hline
grep
& 1135.00
& 20.91
\\
gzip
& 433.33
& 5.25
\\
flex
& 978.63
& 13.15
\\
sed
& 326.37
& 5.18
\\ \hline
\end{tabular}
\label{tab:times}
\end{table}

\begin{figure*}[t]
  \centering
  \subfloat[grep]
  {
    \includegraphics[width=2.5in]{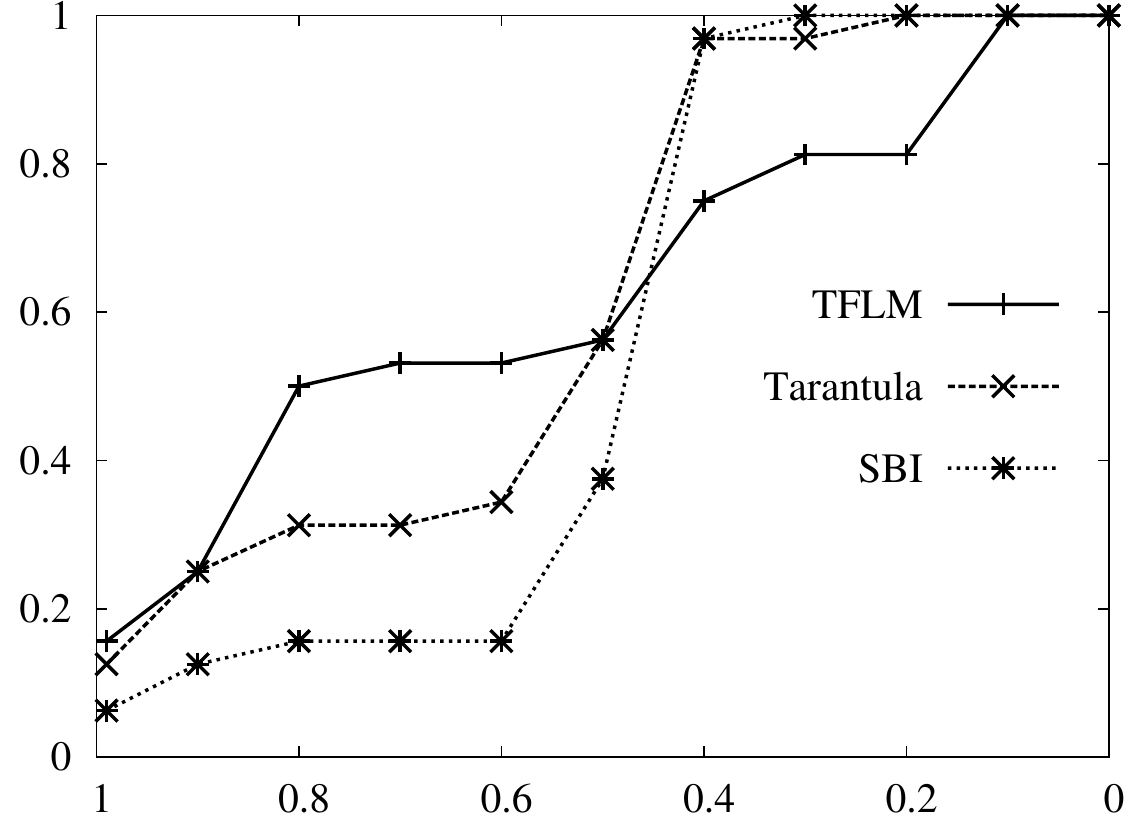}
  }
  \subfloat[gzip]
  {
    \includegraphics[width=2.5in]{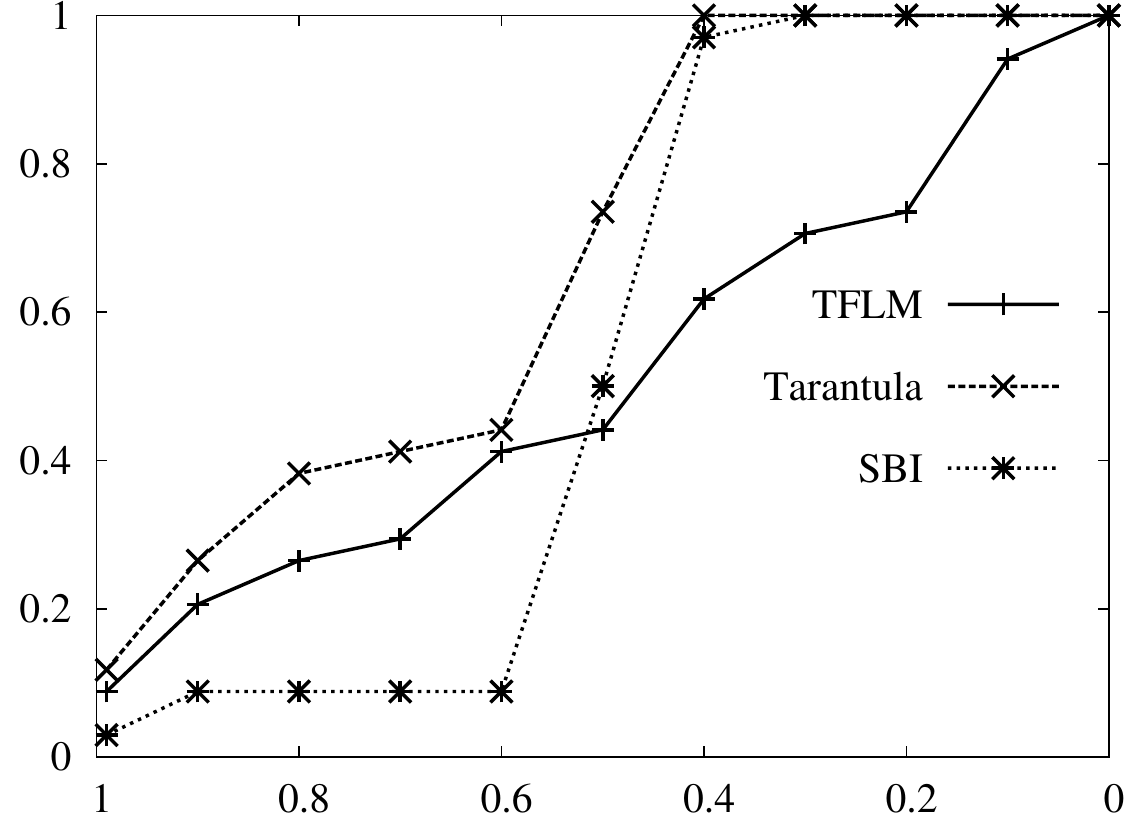}
  } \\
  \subfloat[flex]
  {
    \includegraphics[width=2.5in]{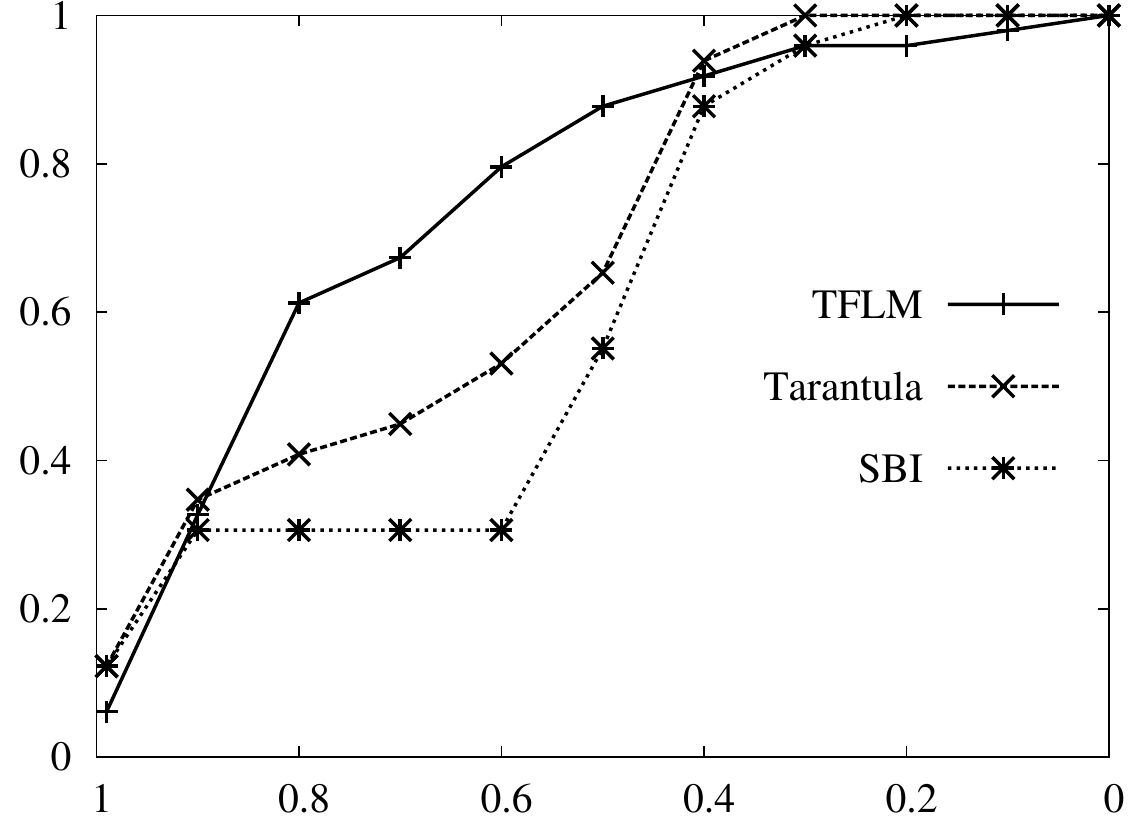}
  }
  \subfloat[sed]
  {
    \includegraphics[width=2.5in]{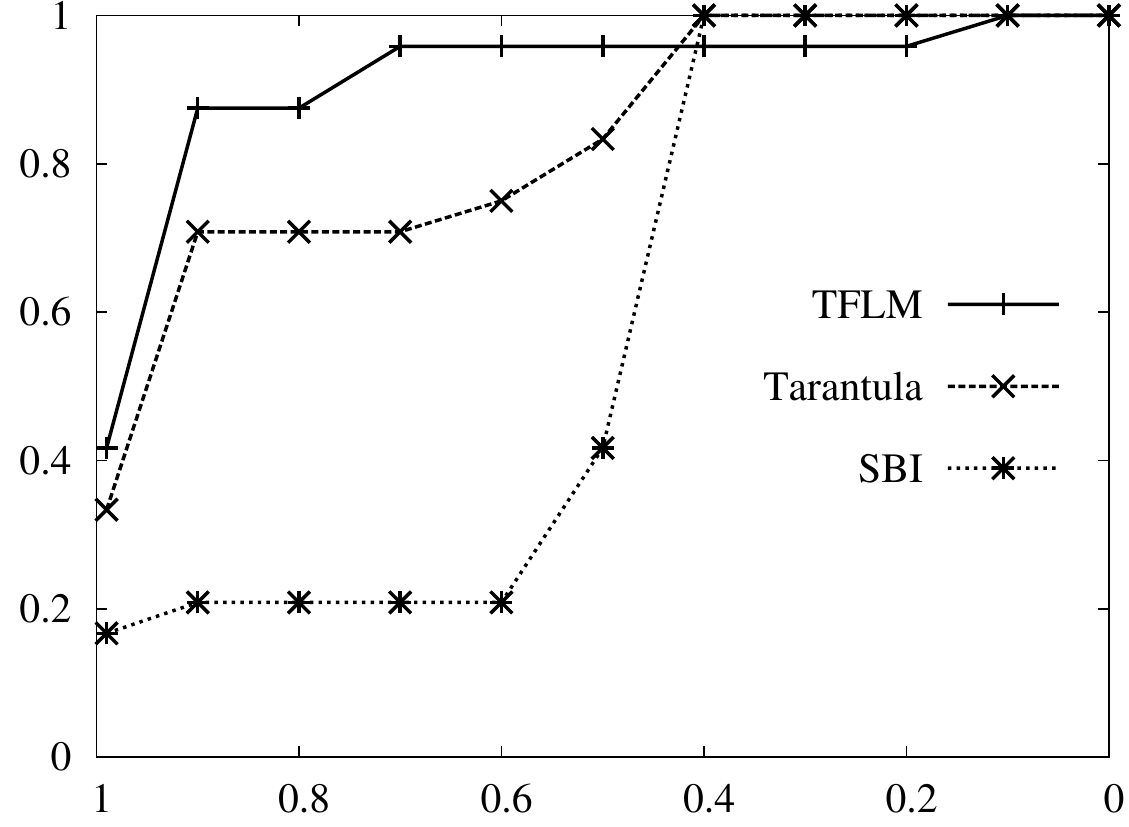}
  }
  \caption{The horizontal axis is the fraction of lines skipped (FS), and the
  vertical axis is the fraction of runs for which the FS score equalled or
  exceeded the x-axis value.}
  \label{fig:results}
\end{figure*}

%

We implemented TFLMs for a simplified version of the C grammar with 23
non-terminal symbols, ranging from compound statements like {\tt if} and {\tt
while} to atomic single-line statements such as assignments and {\tt break} and
{\tt continue} statements.
Each symbol has a $\textit{buggy}$ attribute, and a $\textit{suspiciousness}$ attribute, which is the
{\sc Tarantula} score of the corresponding line. (For AST nodes that correspond
to multiple lines in the original code, we use the highest {\sc Tarantula}
score among all lines). As described in the previous section, the
attributes are modeled as independent univariates during EM ($\textit{buggy}$ as a
Bernoulli distribution, and $\textit{suspiciousness}$ as a Gaussian), and then via a
logistic regression model within each subclass. The model predicts the
$\textit{buggy}$ attribute, using the {\sc Tarantula} score and a bias term as features.
We use the {\sc Scikit-learn} \cite{scikit-learn} implementation of logistic
regression, with the {\tt class\_weight=`auto'} parameter, to compensate for the
sparsity of the buggy lines relative to bug-free lines. For TFLMs, we ran hard
EM for 100 iterations.
For each subject program, TFLMs were learned via cross-validation,
training on all versions of the program except the one being evaluated. The
number of latent subclasses was also chosen via cross-validation, from the
range $[1, 4]$.

The output of a fault localization system is a ranking of the lines of code from
most to least suspicious. For TFLMs, we ranked the lines by predicted
probability that $\textit{buggy}=1$. (Each line in the original program is modeled by the
finest-grained AST node that encloses it.)
The evaluation metric was the `fraction skipped' (FS) score, i.e.\ the fraction
of executable lines ranked below the highest-ranked buggy line.
Despite its limitations \cite{parnin&oso11}, this is a widely-used metric for fault
localization \cite{jones&harrold05,abreu&al09}.
%

\subsection{Results}
Results of our experiments are displayed in tables~\ref{tab:results}
and \ref{tab:compare}, and figure~\ref{fig:results}.
TFLMs outperform {\sc Tarantula} and SBI on three of the four subjects, isolating
the majority of bugs more effectively, and earning a higher average FS score.
However, TFLMs
perform poorly on the {\tt gzip} domain. This demonstrates the main threat
to the validity of our method: machine learning algorithms operate under the
assumption that the test data is drawn from a similar distribution to the
training data.
If the bugs occur in different contexts in the training and test datasets (as in
{\tt gzip}), learning-based methods may perform worse than methods that try to
localize each program independently. This risk is particularly great when the
learning from a small corpus of buggy programs.

However, in three of the four subjects in our experiment, the training and test
distributions are sufficiently similar to allow useful generalization, resulting
in improved fault localization performance.
TFLMs' advantage arises from its ability to localize faults even when the
coverage matrix used by {\sc Tarantula} does not provide useful information
(e.g.\ when the tests are not sufficiently discriminative). TFLMs combine the
coverage-based information used by {\sc Tarantula} with learned bug
probabilities for different symbols, in different contexts. Context sensitivity
is captured via latent subclass assignments for each symbol.

As seen in table~\ref{tab:times}, our unoptimized Python implementation predicts
bug probabilities in a few seconds for programs a few thousand lines in length.
An optimized implementation may be able to make predictions at interactive
speeds; this makes TFLMs a practical choice of inference engine for a
debugging tool in a software development environment. Learning TFLMs can take
several minutes, but note that the model can be trained offline, either from
previous versions of the software being developed (as in our experiments), or
from other related software projects expected to have a similar bug distribution
(e.g.\ projects of a similar scale, written in the same language).

\FloatBarrier
\section{Conclusions}
This paper presented TFLMs, a probabilistic model for fault localization that
can be learned from a corpus of buggy programs. This allows the model to
generalize from previously seen bugs to more accurately localize faults in a new
context. 
TFLMs can also incorporate the output of other fault-localization systems as
features in the probabilistic model, with a learned weight that depends on
the context. TFLMs take advantage of recent advances in tractable probabilistic
models to ensure that the fault location probabilities can be inferred
efficiently even as the size of the program grows.
In our experiments, a TFLM trained with {\sc Tarantula} as a feature localized
bugs more effectively than {\sc Tarantula} or SBI alone, on three of the four
subject programs. 

In this work, we used TFLMs to generalize across sequential versions of a single
program. Given adequate training data, TFLMs could also be used to generalize
across more distantly-related programs. The success of this approach relies on
the assumption that there is some regularity in software faults, i.e.\ the same
kinds of errors occur repeatedly in unrelated software projects, with sufficient
regularity that a machine learning algorithm can generalize over these programs.
Testing this assumption is a direction for future work.

Another direction for future work is extending TFLMs with additional
sources of information, such as including multiple fault localization systems,
and richer program features derived from static or dynamic analysis (e.g.
invariants \cite{hangal&lam02,brun&ernst04}).
TFLM-like models may also be applicable to debugging methods that use path
profiling \cite{chilimbi&al09}, giving the user more contextual
information about the bug, rather than just a ranked list of statements.
The recent developments in tractable probabilistic models may also enable
advances in other software engineering problems, such as fault correction, code
completion, and program synthesis.

\section{Acknowledgments}
This research was partly funded by ARO grant W911NF-08-1-0242,
ONR grants N00014-13-1-0720 and N00014-12-1-0312, and AFRL contract
FA8750-13-2-0019. The views and conclusions contained in this
document are those of the authors and should not be interpreted as
necessarily representing the official policies, either expressed or
implied, of ARO, ONR, AFRL, or the United States Government.
We thank Michael Ernst and Sai Zhang for helpful discussions.

\bibliographystyle{aaai}
{
\bibliography{main}
}
\end{document}